\documentclass[twocolumn,nofootinbib,floats,amsmath,amssymb]{revtex4}
\usepackage{graphicx}
\usepackage{amsmath}
\usepackage{float}
\usepackage{amssymb}
\usepackage[all]{xy}
\usepackage{amsfonts}
\usepackage{dcolumn}
\usepackage{bm}
\usepackage{xcolor}

\begin{document}

\title{Reconstructing Mimetic Cosmology}
 
\author{V\'ictor H. C\'ardenas$^1$}
\email{victor.cardenas@uv.cl}

\author{Miguel Cruz$^2$}
\email{miguelcruz02@uv.mx}

\author{Samuel Lepe$^3$}
\email{samuel.lepe@pucv.cl}

\author{Patricio Salgado$^4$}
\email{patsalgado@unap.cl}

\affiliation{$^1$Instituto de F\'{\i}sica y Astronom\'ia, Universidad de Valpara\'iso, Gran Breta\~na 1111, Valpara\'iso, Chile\\
$^2$Facultad de F\'{\i}sica, Universidad Veracruzana 91000, Xalapa, Veracruz, M\'exico\\
$^3$Instituto de F\'{\i}sica, Facultad de Ciencias, Pontificia Universidad Cat\'olica de Valpara\'\i so, Avenida Brasil 2950, Valpara\'iso, Chile\\
$^4$Instituto de Ciencias Exactas y Naturales, Facultad de Ciencias Universidad Arturo Prat, Avenida Arturo Prat 2120, Iquique, Chile}

\date{\today}

\begin{abstract}
We explore the mimetic gravity formulation with the inclusion of a scalar field potential namely, $V(\phi)$. However, we are not considering any {\it a priori} specific form this term. By means of the Chevallier-Polarski-Linder parametrization for the parameter state of the fluid we can construct an explicit function for such potential in terms of the cosmological redshift and obtain analytical solutions in the mimetic gravity approach. We revise some cosmological implications of these results and additionally we perform a numerical reconstruction for the potential $V$ as a function of the mimetic scalar field, $\phi$.
\end{abstract}

%\pacs{98.80.Cq}

\maketitle

%%%%%%%%%%%%%%%%%%%%%%%%%%%%%%%%%%
\section{Introduction} 
%%%%%%%%%%%%%%%%%%%%%%%%%%%%%%%%%%%
The evidences collected in the era of high energy physics experiments make us confident with respect to the existence of scalar fields in nature \cite{cms}. A seminal idea considering the use of a scalar field can be found in \cite{dicke} and the main motivation to do this was based on the incorporation of the Mach's principle into Einstein's theory of General Relativity (GR), this was the origin of the well-known scalar-tensor models. An interesting description for scalar-tensor theories including more general couplings between gravity and the scalar field was provided by Horndeski in Ref. \cite{horndeski}, and consisted in the emulation of Lovelock gravity \cite{lovelock} at dynamical level: the equations of motion governing the physical degrees of freedom are at most of second order. This characteristic keeps the theory free from certain type of instabilities and it is worthwhile to mention that the Horndeski formulation is the most general scalar-tensor framework preserving second order dynamics. In a more recent context, the {\it Galileon} field (see for instance \cite{galileon}), gained the community's attention on the role of scalar fields in cosmology. In addition to some nice properties inherent to this scalar field, its consequences at cosmological level were relevant, for example; the accelerated cosmic expansion with no need of exotic components, but also an interesting feature of this scalar field is its second order dynamics (and its formulation free from certain type of ghosts) together with its geometric origin as the bending modes of the brane in the scheme of extra dimensions for the Universe\footnote{Other modified theories of gravity as the $f(R)$ theories can also originate a scalar field geometrically \cite{soti}. See also the Ref. \cite{unified}, where it was established that the $f(R)$ gravity framework provides an unified scenario to describe the transition from the early Universe to late times evolution. In Ref. \cite{nut} some generalizations for the $f(R)$ gravity were considered. For example, $f(G)$, being $G$ the Gauss-Bonnet invariant.}.\\ 

Nowadays we can find the consideration of scalar fields in a wide range of physical phenomena and with several purposes, but the use of scalar fields has found a privileged place within cosmology. Our existence itself could be due to a primordial scalar field called {\it inflaton}. This field was the responsible of driving our primitive Universe into a super-accelerated phase which, after a very short time interval, ended with the decay of the inflaton into the particles of the standard model \cite{inflation}. The switching off mechanism for the inflaton still remains unsolved but some interesting proposals can be found in the literature see Ref. \cite{warm}, for instance. The inflationary period of our early Universe is a fundamental ingredient in the current understanding we have about the observable Universe.\\

Scalar-tensor models have proven to be a very prolific theoretical laboratory. It has been shown that in compact objects such as black holes or neutron stars the solutions of GR become unstable when a trivial scalar field is considered, this process is currently known as {\it spontaneous scalarization} \cite{prl}. However, the consideration of more general couplings between the metric and the scalar field can lead to stable configurations \cite{soti2, charmousis}; providing a new wide of solutions for such objects with deviations beyond GR that passes all the solar system tests.\\

It is a known fact that the conformal transformations, can help to shed light on a vast class of scalar-tensor theories\footnote{For instance, in Ref. \cite{faraoni} is discussed that the gravitational lensing generated by scalar-tensor gravitational waves is stronger in the Jordan frame than in the Einstein frame; with the recent detection of gravitational waves we could start to get some hints about the equivalence or inequivalence between both frames, which has been a subject of controversy among cosmologists for several years. For a review on conformal transformations see also \cite{sokoloswki}. The conformal transformations with multiple scalar fields can be found in \cite{kaiser}.}. However, when trying to apply the same reasoning to a more general class of scalar-tensor theories, such as those included in the Horndeski action, it is found that conformal transformations do not work as in the standard case because of the kinetic dependence in the free parameters of the Horndeski theory. The conformal transformations are metric transformations consisting on a point-dependent re-scaling of the metric tensor. For dynamical reasons it is usual to assume that the conformal factor has a functional dependence on the scalar field appearing in the theory by means of its first derivatives, i.e., 
\begin{eqnarray}
\tilde{g}_{\mu \nu } \longrightarrow g_{\mu \nu } &=& \Omega ^{2}\left( \phi
,X\right) \tilde{g}_{\mu \nu }  \label{eq:1} \\
X &=&\tilde{g}^{\mu \nu }\partial _{\mu }\phi \partial _{\nu }\phi.  \nonumber
\end{eqnarray}
Here $X$ is the simplest coordinate-invariant we can think of using only the metric tensor and the scalar field. In Ref. \cite{fla} was found that the so-called ``standard scalar-tensor theory'' is closed under the conformal transformation $g_{\mu \nu }=\Omega^{2}\left( \phi \right) \tilde{g}_{\mu \nu }$. The natural question is
whether this is also true for the extended conformal transformation (\ref{eq:1}). The answer is negative. The $X$-dependence introduces new terms which cannot be brought back to the canonical form of the standard scalar-tensor theory. The interesting thing about the extended conformal transformations (\ref{eq:1})
is that inside its domain lies the Mimetic Gravity theory \cite{mukhanov1}. These theories arise from the fact that not always the conformal transformations (\ref{eq:1}) are invertible. If this was the case then to the non-invertibility condition corresponds an extra degree of freedom in the theory, and in fact a new physically different theory with respect to the untransformed one. An example of extended conformal transformation giving rise to a Mimetic degree of freedom is given by 
\begin{equation}
\tilde{g}_{\mu \nu }\longrightarrow g_{\mu \nu }=\left( -\tilde{g}^{\alpha \beta
}\partial _{\alpha }\phi \partial _{\beta}\phi \right) \tilde{g}_{\mu \nu }
\label{2}
\end{equation}
that corresponds to $\Omega ^{2}=X^{-1}$. Actually this was the first example of Mimetic gravity \cite{mukhanov1} (see also \cite{izaurieta}).\\

Our aim in this work is to reconstruct a viable Mimetic Gravity with cosmological observations of the late times epoch by means of a Chevallier-Polarski-Linder (CPL) parametrization which takes the parameter state, $\omega$, to be a linear function of the scale factor, namely \cite{cpl, cpl2}
\begin{equation}\label{eqcpl}
    \omega(a) = \omega_{0} +\omega_{a}(1-a),
\end{equation}
where $\omega_{0}$ and $\omega_{a}$ are constants. As we will discuss later, the emergence of dark energy or deviations from the standard $a^{-3}$ behavior in the Mimetic Gravity approach are induced by the presence of a potential for the mimetic field. On the other hand, the constant behavior for dark energy (as a cosmological constant) lacks of physical arguments in any theoretical model, which then opens the possibility to consider alternative proposals such as a dynamical dark energy component or a redshift dependence in the EoS parameter $\omega(z)$ as in the case of the CPL parametrization. As we will see below, the reconstructed Mimetic potential depends only on the values of the constants $\omega_{0}$ and $\omega_{a}$ appearing in the CPL parametrization and varies with the scalar field (made possible from a numerical reconstruction for the shape of $V$) or in terms of redshift; such free parameters have been constrained using several observational data sets. In particular we restrict ourselves to consider only three cases. Other reconstructions for Mimetic Gravity have been considered before but in those approaches the compatibility of the model is established with the inflationary stage, see for instance Ref. \cite{recons1}.\\

The outline of this paper is as follows. In Section \ref{sec:mimetic} we provide some highlights of the Mimetic Gravity formulation with the inclusion of a Lagrange multiplier at the action level. In Section \ref{sec:cpl} we discuss the implementation of the CPL parametrization for the equation of state parameter of the mimetic fluid and from this consideration we construct the potential $V$, which results as a function of the cosmological redshift. We explore some cosmological aspects of this solution and we numerically reconstruct the scalar field potential $V(\phi)$. In Section \ref{sec:final} we write our final comments. We will use $8\pi G=c=k_{B}=1$ units throughout this work.      

%%%%%%%%%%%%%%%%%%%%%%%%%%%%%%%%%%%%%%%%%%%%%%%%%%%%%%%%%%%%%%%
\section{Mimetic gravity}
\label{sec:mimetic}
%%%%%%%%%%%%%%%%%%%%%%%%%%%%%%%%%%%%%%%%%%%%%%%%%%%%%%%%%%%%%%
In Ref. \cite{Beken} was introduced a class of metric transformations, dubbed as {\it disformal transformations}. Bekenstein \cite{Beken} considered gravitational theories supplied by two geometries, one for the gravity sector and the other for the matter sector, such as in Brans-Dicke type theories, where the matter metric is related to the gravity metric by a conformal transformation. In this reference Bekenstein finds that because GR enjoys invariance under diffeomorphisms, one is free to parametrize the metric $g_{\mu \nu}$ in terms of an auxiliary metric $\tilde{g}_{\mu \nu}$ and a scalar field $\phi$. The transformation between these two metrics is known as ``disformal transformation'' and is given by 
\begin{equation}
g_{\mu \nu }=A\left( \phi ,X\right) \tilde{g}_{\mu \nu }+B\left( \phi
,X\right) \partial _{\mu }\phi \partial _{\nu }\phi ,  \label{3}
\end{equation}%
where $X=\tilde{g}^{\mu \nu}\partial _{\mu}\phi \partial_{\nu}\phi$. Here the functions $A$ and $B$ are scalar parameters, called the conformal and disformal factors, respectively. For $B=0$ the previous expression reduces to the extended conformal transformation. In general, the functions $A\left( \phi ,X\right)$ and $B\left(\phi ,X\right)$ are arbitrary, with $A \neq 0$. In Ref. \cite{deruel} was shown that, provided the transformation is invertible, the equations of motion for the theory, obtained by the variation of the action with respect to $\tilde{g}_{\mu \nu}$ and $\phi$, reduce to those obtained by varying with respect to the metric $g_{\mu \nu}$. In the same reference \cite{deruel} and in \cite{mimet1} was shown that if the equations of motion are constrained with $\tilde{g}_{\mu \nu}$ and $\phi$, then the equations of motion are given by a system of equations whose determinant can be zero. If this is the case, then the disformal transformation given by (\ref{3}) is non-invertible or singular and we find extra degrees of freedom which result in equations of motion that differs from those of GR \cite{deruel}. The parametrization of reference \cite{mukhanov1} defining mimetic gravity can be identified with a singular disformal transformation, with $A=X$ and $B=0$ in Eq. (\ref{3}). This singularity of the disformal transformation has as consequence the existence of extra degrees of freedom in the system; explaining the origin of the extra degree of freedom in mimetic gravity.\\

In Ref. \cite{mimet3} was shown that the two approaches towards mimetic gravity, namely, singular disformal transformation (\ref{3}) and the so called Lagrange multiplier formulation are equivalent. The idea of the authors of Ref. \cite{mukhanov1} is to isolate the conformal degree of freedom of gravity by introducing a parametrization of the physical metric $g_{\mu \nu}$ in terms of an auxiliary metric $\tilde{g}_{\mu \nu}$ and a scalar field $\phi$, dubbed {\it mimetic field}, as follows
\begin{equation}
g_{\mu \nu }=\left( -\tilde{g}^{\alpha \beta}\partial _{\alpha}\phi \partial _{\beta}\phi \right) \tilde{g}_{\mu \nu }.  \label{2'}
\end{equation}
From (\ref{2'}) it is clear that, in such a way, the physical metric is invariant under conformal transformations of the type, $\tilde{g}_{\mu \nu }\longrightarrow \Omega (t,x)^{2}\tilde{g}_{\mu \nu
}$, for the auxiliary metric; being $\Omega (t,x)$ a function of the space-time coordinates. It is also clear that, as a consistency condition, the mimetic field satisfies the following constraint
\begin{equation}
g^{\mu \nu }\partial _{\mu }\phi \partial _{\nu }\phi =-1.  \label{eq:variation}
\end{equation}
Thus, the gravitational action, taking into account the reparametrization given by (\ref{2'}) now takes the form
\begin{equation}
S=\int_{M}d^{4}x\sqrt{-g\left( \tilde{g}_{\mu \nu },\phi \right) }\left\lbrace
R\left( \tilde{g}_{\mu \nu },\phi \right) +\mathcal{L}_{m}\right\rbrace,  \label{9}
\end{equation}%
where $M$ is the spacetime manifold, $R=R\left( \tilde{g}_{\mu \nu },\phi \right)$ is the Ricci scalar, $\mathcal{L}_{m}$ is the matter Lagrangian and $g=g\left( \tilde{g}_{\mu \nu },\phi \right)$ is the determinant of the physical metric. By varying the action with respect to the physical metric one obtains the
equations for the gravitational field \cite{mukhanov1} 
\begin{equation}
G_{\mu \nu }-T_{\mu \nu }-\left( G-T\right) \left( \partial _{\mu }\phi
\right) \left( \partial _{\nu }\phi \right) =0,  \label{10}
\end{equation}%
where $G_{\mu \nu }$ and $T_{\mu \nu }$ are the Einstein tensor and the matter energy-momentum tensor, while $G$ and $T$ represent the traces of such tensors, respectively. Notice that the mimetic field contributes to the right hand side of Einstein's equation through the additional energy-momentum tensor component
\begin{equation}
\tilde{T}_{\mu \nu }=-\left( G-T\right) \left( \partial _{\mu }\phi \right)
\left( \partial _{\nu }\phi \right).  \label{11}
\end{equation}
In reference \cite{mimet4} was considered an alternative but equivalent formulation for mimetic gravity. The equations of motion obtained from the action written in terms of the auxiliary metric $\tilde{g}_{\mu \nu}$ are equivalent to those that one would conventionally obtain from the action expressed in terms of the physical metric with the imposition of an additional constraint on the mimetic field. In the formulation of the reference \cite{mimet4} the mimetic constraint given by (\ref{eq:variation}) can actually be implemented at the level of the action by using a Lagrange multiplier. That is, the action for mimetic gravity (\ref{9}) can be written as
\begin{eqnarray}
    S &=& \int_{M} d^{4}x\sqrt{-g}\left\lbrace \frac{R(g)}{2}-\frac{\lambda}{2}\left[g^{\mu \nu}(\partial_{\mu}\phi)(\partial_{\nu} \phi) -1\right] +  \right. \nonumber \\
     &+& \left. \mathcal{L}_{m} - V(\phi) \right\rbrace,
    \label{eq:modified}
\end{eqnarray}
where we have considered a potential for the mimetic field. The variation of the action (\ref{eq:modified}) with respect to the physical metric, $g_{\mu \nu}$, leads to the following equations of motion \cite{mukhanov2}
\begin{equation}
    G_{\mu \nu} = T_{\mu \nu}+\lambda \partial_{\mu}\phi \partial_{\nu} \phi - g_{\mu \nu}\left[\frac{\lambda}{2}(\partial^{\alpha}\phi \partial_{\alpha} \phi -1) + V(\phi)\right],
    \label{eq:modified2}
\end{equation}
On the other hand, the variation with respect to the Lagrange multiplier field $\lambda$ provides the condition (\ref{eq:variation}) while the variation with respect to the mimetic field $\phi$ reads
\begin{equation}
    \nabla^{\mu}(\lambda \partial_{\mu}\phi) - \frac{dV(\phi)}{d \phi} = 0,
    \label{eq:gordon}
\end{equation}
which corresponds to a generalization of the Klein-Gordon equation. If we take the trace of Eq. (\ref{eq:modified2}) and consider the condition given in (\ref{eq:variation}) we can obtain an explicit expression for the Lagrange multiplier
\begin{equation}
    \lambda = G - T + 4V,
\end{equation}
substituting the previous expression for the Lagrange multiplier in (\ref{eq:modified2}) and considering $V = 0$ together with the condition (\ref{eq:variation}), we recover the equation (\ref{10}).\\ 

Now, we consider a flat FLRW metric of the form $ds^{2} = -dt^{2} + a^{2}(t)(dx^{2}+dy^{2}+dz^{2})$, where $a(t)$ is the scale factor. We must take into account that in order to preserve the homogeneity and isotropy of spacetime we must have for the scalar field, $\phi = \phi(t)$. From Eq. (\ref{eq:variation}) we obtain, $\dot{\phi}^{2} = 1$, which leads immediately to $\phi(t) = t$. For convenience we have chosen a null value for the integration constant. The Friedmann constraint and acceleration equation follows from (\ref{eq:modified2}) and can be written as
\begin{align}
    & 3H^{2} = \rho + \lambda + V, \label{eq:friedmann}\\
    & 2\dot{H} + 3H^{2} = V - p \Rightarrow \dot{H} + H^{2} = - \frac{1}{6}(\rho + 3p + \lambda - 2V),
    \label{eq:accel}
\end{align}
where the energy momentum tensor of the matter content is modeled by a perfect fluid, being $\rho$ and $p$ its density and pressure, respectively. In our description, we will consider a barotropic fluid, therefore the equation state to consider will have the form, $p_{\mathrm{x}} = \omega_{\mathrm{x}}\rho_{\mathrm{x}}$. Note that the term $\lambda + V$ in the Friedmann constraint (\ref{eq:friedmann}) is the contribution coming from mimetic gravity to the energy density of the fluid, for simplicity we will call it $\rho_{\mathrm{MG}}$. From the acceleration equation (\ref{eq:accel}) we can identify the pressure added by mimetic gravity, $p_{\mathrm{MG}} = -V$. Using the Eqs. (\ref{eq:friedmann}) and (\ref{eq:accel}) we can obtain for both energy densities
\begin{equation}
    \dot{\rho} + 3H\rho\left(1+\omega\right) + \dot{\rho}_{\mathrm{MG}} + 3H\rho_{\mathrm{MG}}\left(1+\omega_{\mathrm{MG}}\right) = 0,
    \label{eq:continuity}
\end{equation}
for barotropic fluids. If we consider the continuity for the matter sector, $\nabla^{\mu}T_{\mu \nu} = 0$, and a barotropic equation of state, we obtain the standard form $\dot{\rho} + 3H\rho\left(1+\omega\right) = 0$. On the other hand, from the identifications $\rho_{\mathrm{MG}} = \lambda + V$ and $p_{\mathrm{MG}} = -V$, we can verify with the use of Eqs. (\ref{eq:friedmann}) and (\ref{eq:accel}) that $\dot{\rho}_{\mathrm{MG}} + 3H(\rho_{\mathrm{MG}}+p_{\mathrm{MG}}) = 0$, i.e., the energy density associated to the Mimetic Gravity sector is also conserved \cite{rinaldi}. From the acceleration equation (\ref{eq:accel}) we can obtain for a barotropic fluid description the deceleration parameter after a straightforward calculation, yielding
\begin{equation}
    q = \frac{1}{2}+\frac{3}{2}\left(\frac{\omega \rho + \omega_{\mathrm{MG}}\rho_{\mathrm{MG}}}{\rho + \rho_{\mathrm{MG}}}\right).
\end{equation}
Using the fact that $\lambda = \rho_{\mathrm{MG}} - V$, the equation (\ref{eq:gordon}) can be integrated to obtain
\begin{eqnarray}
\rho_{\mathrm{MG}} &=& \frac{c}{a^{3}} + \frac{3}{a^{3}}\int^{t}_{t_{0}}a^{3}(t')H(t')V(t')dt' \label{eq:densitya}\\ &=& \frac{c}{a^{3}} + \frac{3}{a^{3}}\int a^{2}V da,
\label{eq:densityb}
\end{eqnarray}
being $c$ an integration constant. It is worthy to mention that for $V=0$, the mimetic scalar field resembles the dark matter behavior since $\rho_{\mathrm{MG}} = G - T \propto a^{-3}$ and $p_{\mathrm{MG}} = 0$. Typically one can have explicit solutions for $\rho_{\mathrm{MG}}$ until a specific Ansatz for $V(\phi)$ is assumed, an extensive literature can be found on this topic following a route of this kind.

%%%%%%%%%%%%%%%%%%%%%%%%%%%%%%%%%%%%%%%%%%%%%%%%%%%%
\section{CPL parametrization: reconstructing the potential}
\label{sec:cpl}
%%%%%%%%%%%%%%%%%%%%%%%%%%%%%%%%%%%%%%%%%%%%%%%%%%%%

\begin{figure}[htbp!]
    \centering
    \includegraphics[width=3.0in]{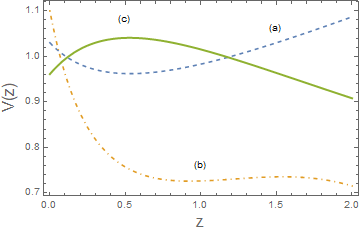}
    \caption{$V(z)$ using the best fit values for $w_0$ and $w_a$. Case (a): $\omega_{0} = -1.03, \ \omega_{a} = 0.26$, case (b) corresponds to $\omega_{0} = -1.1, \ \omega_{a} = 1.41$ of Ref. \cite{datacpl} and case (c) $\omega_{0} = -0.96, \ \omega_{a} = -0.29$ from Planck \cite{planck}.}
    \label{fig:vdzabc}
\end{figure}

Instead of choosing a specific form for the scalar field potential, let us consider the case where the corresponding equation of state parameter for the mimetic fluid is assumed to take the form of the CPL \cite{cpl, cpl2} parametrization as given in Eq. (\ref{eqcpl}) this time written in terms of redshift
\begin{equation}
    \omega_{\mathrm{MG}} = \omega_{0}+\omega_{a}\frac{z}{1+z},
\end{equation}
where $\omega_{0}$ and $\omega_{a}$ are constants which can have the following natural physical interpretation: the present time value of the parameter state and its overall time evolution, respectively. This parametrization has been widely used in the literature and has become the standard to study the cosmic evolution of the EoS parameter \cite{planck}. It describes a soft evolution with a constant bounded value $\omega_{0} + \omega_{a}$ for the parameter state at early times towards a present day value of $\omega_{0}$, so we can safely describe the entire evolution of the universe from $z=\infty$ to the present day at $z=0$, although we cannot extend the study too far into the future because the parametrization breaks at $z=-1$. It is worthy to mention that other possibilities compatible with cosmological observations for $\omega(z)$ can be found, see for instance Ref. \cite{otheromega}. Therefore from the previous expression and the conservation condition for $\rho_{\mathrm{MG}}$, the energy density evolves as a function of the redshift as follows
\begin{equation}
    \rho_{\mathrm{MG}}(z) = \rho_{0}(1+z)^{3(1+\omega_{0}+\omega_{a})}\exp \left(-3\omega_{a}\frac{z}{1+z} \right),
    \label{eq:densitycpl}
\end{equation}
where we have considered the standard relation between the scale factor and the redshift, $1+z = a^{-1}$. On the other hand, if we equate the previous expression for the energy density with the Eq. (\ref{eq:densityb}) one gets an explicit form of the potential given as a function of the redshift 
\begin{eqnarray}
    V(z) &=& - \rho_{0}\left[\omega_{0}+(\omega_{0}+\omega_{a})z \right](1+z)^{2+3(\omega_{0}+\omega_{a})} \times \nonumber \\
    &\times & \exp \left(-3\omega_{a}\frac{z}{1+z} \right).
    \label{eq:potential}
\end{eqnarray}
Note that at present time ($z=0$) the potential takes the constant value $V=-\rho_{0}\omega_{0}$ and at the far future, i.e., in the limit $z\rightarrow -1$ we have $V(z \rightarrow -1) \rightarrow 0$ only if $\omega_{a} < 0$. We display the potential $V(z)$ in Fig. (\ref{fig:vdzabc}).

%%%%%%%%%%%%%%%%%%%%%%%%%%%%%%%%%%%%%%%%%%%
%%%%%%%%%%%%%%%%%%%%%%%%%%%%%%%%%%%%%%%%%%%%5

Notice that for case (a) the potential has a minimum around $z \simeq 0.6$ and start to increase to the past (large $z$). In the case (b) the potential decrease rapidly as redshift increase and there seems to reach a minimum around $z \simeq 0.7$ and after a small increase reaching a maximum around $z \simeq 1.5$. In the case (c), the potential does not have a minimum and decrease monotonically as redshift increases. It is worthy to mention that the constant values $\omega_{0}$ and $\omega_{a}$ for the cases (a) and (b) found in Ref. \cite{datacpl} were constrained using SN Ia and BAO data by performing a comparison between the CPL parametrization and a fiducial step-like dark energy model. This latter dark energy model also consists on a parametrization for the parameter state. However, in this case we have to deal with four free parameters instead two. An advantage of this parametrization is its analytical integrability; besides, the asymptotic values for the parameter state before and after the transition are decoupled. In general, these models also have the property of characterizing by election a fast or slow transition between the past and future evolution of the Universe.\\

From the identification, $\lambda = \rho_{\mathrm{MG}} - V$, we can have also an explicit expression for the Lagrange multiplier
\begin{eqnarray}
    \lambda(z) &=& \rho_{0}(1+z)^{2+3(\omega_{0}+\omega_{a})}\left[1+\omega_{0}+(1+\omega_{0}+\omega_{a})z \right]\times \nonumber \\ &\times & \exp \left(-3\omega_{a}\frac{z}{1+z} \right).
    \label{eq:multiplier}
\end{eqnarray}
At present time the Lagrange multiplier is only a constant given by $\lambda_{0} = \rho_{0}(1+\omega_{0})$, which lies in the interval $\rho_{0}[-0.1,0.04]$ in the cases (a), (b) and (c) commented before for the pair of values $\omega_{0}$ and $\omega_{a}$. It is worthwhile to mention that in the absence of scalar field potential we have, $\lambda = \rho_{\mathrm{MG}}$, and from Eq. (\ref{eq:densityb}) we observe that the Lagrange multiplier can be used to model standard dark matter since it decays as $a^{-3}$. Then, the responsible of having a late time accelerated expansion for the Universe is $V(\phi)$ or $V(t)$, because for $V \neq 0$ the $a^{-3}$ behavior for the Lagrange multiplier is modified as can be seen in expression (\ref{eq:multiplier}). This deviation from the $a^{-3}$ behavior is characteristic of models in which interaction is allowed between dark matter and dark energy, represented in our case by $\lambda$ and $\phi$, respectively \cite{interact, de}. Some results show that the interaction of the mimetic field with particles of the standard model such as photons and baryons lead to predictions that do not contradict some observable phenomena of our Universe. In fact, this latter scenario provides an interesting framework from the particle physics point of view since the CPT symmetry could be broken spontaneously \cite{photons}. On the other hand, according to the behavior of $V(\phi)$ we will have matter or dark energy domination; for instance if we have, $V(\phi) \rightarrow 0$, then the interaction turns off and the dark matter behavior $a^{-3}$ dominates. The shape of the potential $V(z)$ determines the behavior of the energy density $\rho_{\mathrm{MG}}(z)$ by means of Eq. (\ref{eq:densityb}). For the cases (a) and (b) the potential grows from present time ($z=0$) to far future ($z=-1$) while in the case (c) it decreases because we have $\omega_{a} < 0$. Therefore the corresponding density for each case behaves similarly as $V(z)$.\\ 

\begin{figure}[htbp!]
\centering
\includegraphics[width=7.8cm,height=6.5cm]{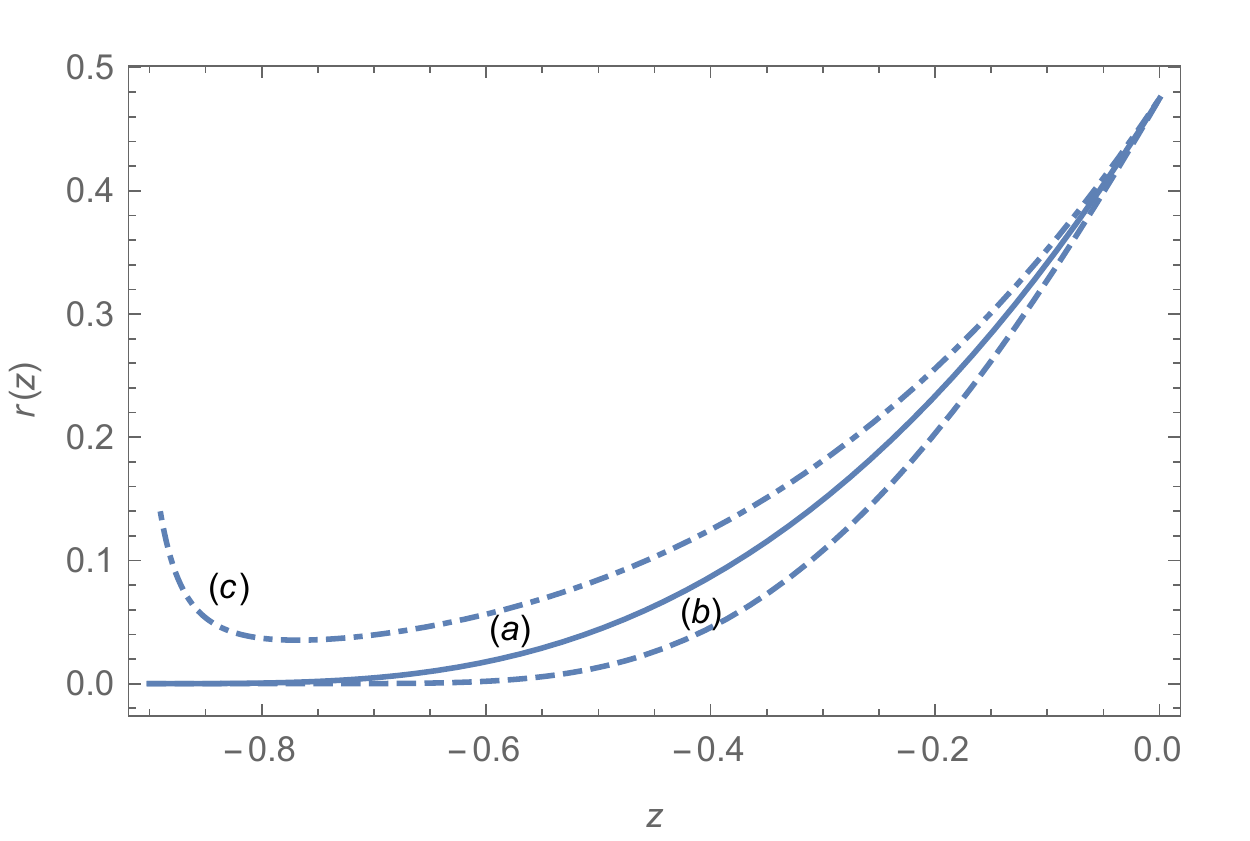}   
\caption{Coincidence parameter, $r(z)$.}
\label{fig:coincidence}
\end{figure}

If we assume a typical dark matter contribution sector together with the energy density given in (\ref{eq:densitycpl}) for the mimetic gravity sector and the Friedmann constraint (\ref{eq:friedmann}), we obtain for the normalized Hubble parameter
\begin{align}
& E^{2}(z) := \frac{H^{2}(z)}{H^{2}_{0}} = \Omega_{\mathrm{m,0}}(1+z)^{3} + \nonumber \\
&    + \Omega_{\mathrm{MG,0}}(1+z)^{3(1+\omega_{0}+\omega_{a})}\exp \left(-3\omega_{a}\frac{z}{1+z} \right),
\label{eq:hubblecpl}
\end{align}
where we have considered the usual definition for the density parameters, $\Omega_{\mathrm{x,0}} := \rho_{\mathrm{x,0}}/3H^{2}_{0}$. Note that in this case we must have the normalization condition $\Omega_{\mathrm{m,0}} + \Omega_{\mathrm{MG,0}} = 1$. A remark about the previous expression for the normalized Hubble parameter is in order; depending on the value of the constant $\omega_{a}$, the $E(z)$ function could diverge at $z=-1$, this corresponds to a little big rip singularity. This scenario for dark energy represents a viable alternative to the $\Lambda$CDM model since the future singularity is avoided for a finite value of the cosmological redshift \cite{little}. On the other hand, if we consider a Universe filled with matter and dark energy which is modeled by the mimetic energy density, then we can write the coincidence parameter from Eq. (\ref{eq:hubblecpl}) which is given as the quotient between the dark matter and dark energy densities as follows
\begin{equation}
    r(z) = r_{0}(1+z)^{-3(\omega_{0}+\omega_{a})}\exp \left(3\omega_{a}\frac{z}{1+z} \right),
\end{equation}
where we have defined $r_{0} := \Omega_{\mathrm{m,0}}/(1-\Omega_{\mathrm{m,0}})$. The behavior of the coincidence parameter is shown in Fig. (\ref{fig:coincidence}) with the cases (a), (b) and (c) for $\omega_{0}$ and $\omega_{a}$, as discussed before and $0.315 \pm 0.007$ \cite{planck} for $\Omega_{\mathrm{m,0}}$. For the cases (a) and (b), the coincidence parameter decreases and tends to zero from present time ($z=0$) to the future ($z\rightarrow -1$). However, for the case (c) we have that around $z \simeq -0.8$, this parameter changes its tendency and starts to increase. For this latter case we have a distinct behavior from $\Lambda$CDM model, where $r(z\rightarrow -1) \rightarrow 0$ \cite{coincidence}, therefore the behavior for the coincidence parameter in case (c) is not the desired one. Then under the appropriate election of the parameters involved in the model, the cosmological coincidence problem can be alleviated if Mimetic Gravity models the dark energy content of the Universe with a CPL parametrization for the parameter state. Besides, as can be seen in the plot, from the recent past ($z>0$) to the present time, the coincidence parameter has a decreasing behavior and $r(z \rightarrow \infty) \rightarrow \infty$, for the three cases; this means that at some point of the cosmic evolution the dark energy component becomes dominant over the matter content. It is worthy to mention that at present time the cases (a), (b) and (c) coincide.      

%%%%%%%%%%%%%%%%%%%%%%%%%%%%%%%%%%%%%%%%%%%%%%%%5
%%%%%%%%%%%%%%%%%%%%%%%%%%%%%%%%%%%%%%%%%%%

\begin{figure}[htbp!]
\centering
\includegraphics[width=8.4cm,height=8.0cm]{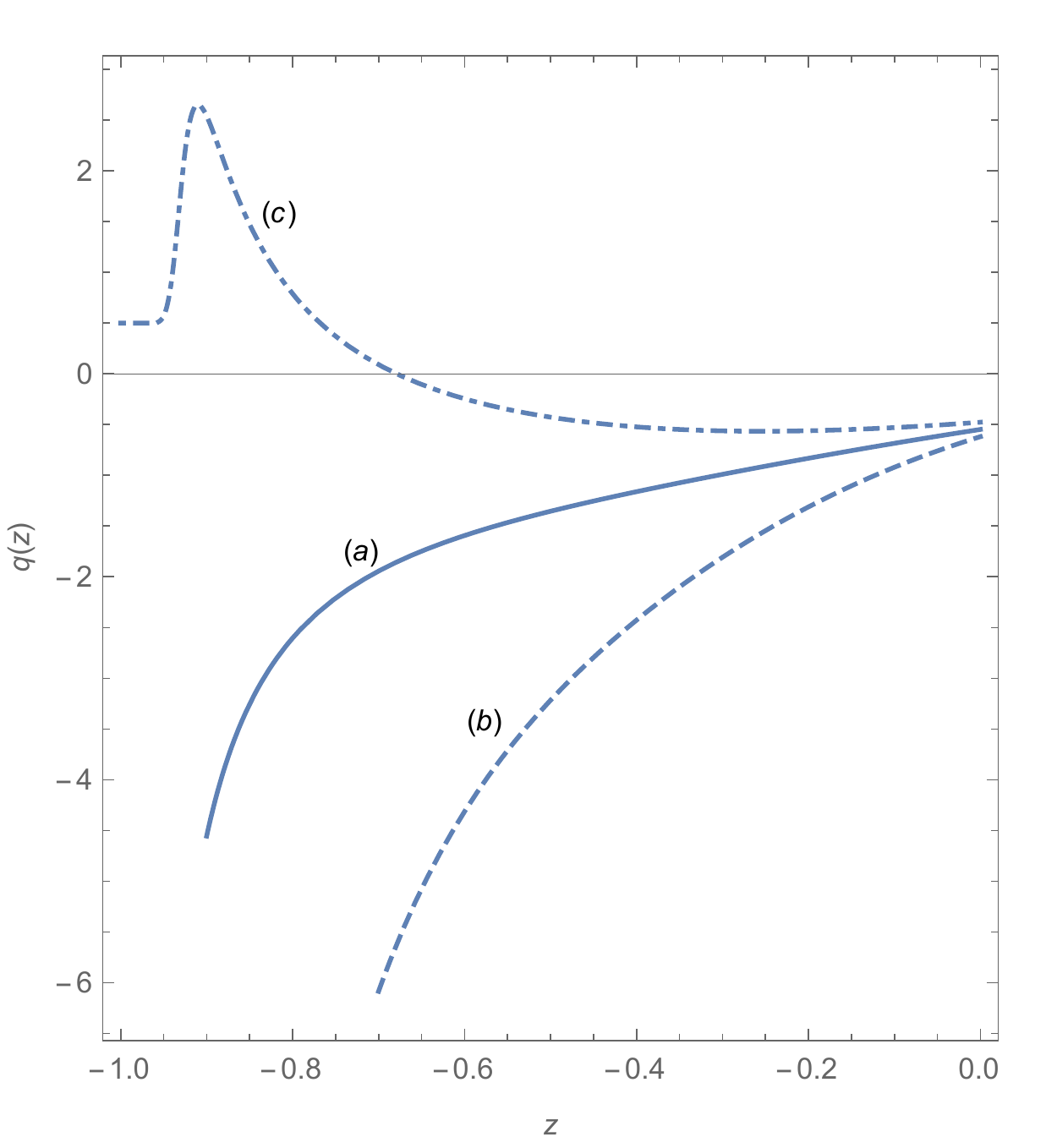}   
\caption{Behavior of the deceleration parameter.}
\label{fig:decelcpl}
\end{figure}

In terms of the normalized Hubble parameter we can write the deceleration parameter as follows
\begin{equation}
    q(z) = - 1 + (1+z)\frac{d \ln E(z)}{dz}.
\end{equation}
In Fig. (\ref{fig:decelcpl}) we depict the previous expression for the deceleration parameter with the Eq. (\ref{eq:hubblecpl}) for $E(z)$. We have used the same values for the constant parameters as was done in the previous plots for the cases (a), (b) and (c). As can be seen, $q(z = 0) < 0$ in the three cases, however only the case (a) and (b) maintain negative values (and increasing) as the Universe expands. At some stage of the cosmic evolution these models, the cases (a) and (b), could mimic the $\Lambda$CDM model ($q=-1$), but as can be seeing as we move to the future they reach a behavior where $q < -1$, which represents a phantom regime. Finally, for the case (c) we can observe that the cosmic evolution moves from a quintessence dark energy behavior to a decelerated expansion, i.e., in this case we can have a Milne Universe characterized by, $q(z)=0$, at some stage of the cosmic evolution. 

%%%%%%%%%%%%%%%%%%%%%%%%%%%%%%%%%%%%%%%%%%%%%%%%5
%%%%%%%%%%%%%%%%%%%%%%%%%%%%%%%%%%%%%%%%%%%

Now we want to reconstruct $V(\phi)$ numerically. For this task, we will use $V(z)$ from (\ref{eq:potential}) and the field $\phi$ is obtained from the Hubble function definition $H = \dot{a}/a$ which implies
\begin{equation}
    dt = \frac{da}{aH} = - \frac{dz}{(1+z)H(z)},
\end{equation}
where we have used that $a=(1+z)^{-1}$. This means that $t(z)$ satisfy the equation
\begin{equation}\label{dtdz}
    \frac{dt}{dz} + \frac{1}{(1+z)H(z)}=0.
\end{equation}
Before use this numerically, let us divide by $t_0 = t(z=0)$ then defining the new variable $\tau = t/t_0$ the equation leads to
\begin{equation}\label{dtaudz}
    \frac{d\tau}{dz} + \frac{1}{t_0H_0}\frac{1}{(1+z)E(z)}=0,
\end{equation}
where now $\tau $ evolve in the range $(0,1)$ and $E(z)$ is defined by the Eq. (\ref{eq:hubblecpl}).
%\begin{equation}
%    E^2(z)=\Omega_m (1+z)^3 + (1-\Omega_m)(1+z)^{3(1+w_0+w_a)}e^{-3w_az/(1+z)}.
%\end{equation}
For the $\Lambda$CDM model the quantity $t_0H_0$ is given by
\begin{equation}
    t_0H_0 = \frac{2}{3\sqrt{\Omega_V}}\ln \frac{1+\sqrt{\Omega_V}}{\sqrt{1+\Omega_V}} \simeq 0.964
    \label{eq:time}
\end{equation}
which the numerical value applies for $\Omega_V =0.7$.\\

\begin{figure}[h!]
    \centering
    \includegraphics[width=3.0in]{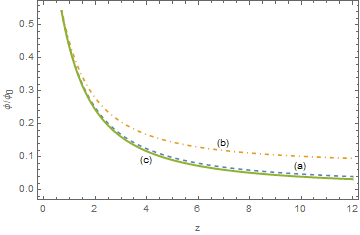}
    \caption{$t(z)$ from Eq.(\ref{dtdz}) using the best fit values for $w_0$ and $w_a$ for each of the three set data discussed in the text.}
    \label{tdzabc}
\end{figure}

We display three realizations of $V(\phi)$ for different pairs of CPL parameters: (a) $w_0=-1.03$, $w_a=0.26$, (b) $w_0=-1.1$, $w_a=1.41$ from \cite{datacpl} and (c) $w_0=-0.961$ and $w_a=-0.29$ from Planck \cite{planck}. Using (\ref{dtaudz}) we can integrate $t(z)$ for each one of our three cases. In Fig. (\ref{tdzabc}) we display the results of the integration.
Notice that we have written $\phi/\phi_0$ in the vertical axis, understanding for $\phi_0$ as the value of the field today.\\

\begin{figure}[h!]
    \centering
    \includegraphics[width=3.0in]{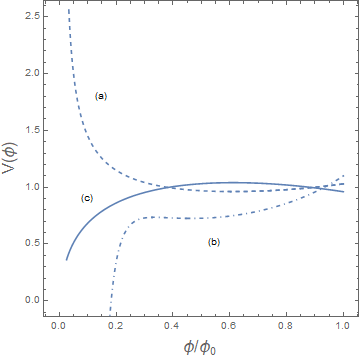}
    \caption{$V(\phi)$ as a function of the field value $\phi$ using the best fit values for $w_0$ and $w_a$ for each of the three set data discussed in the text.}
    \label{vdfiabc}
\end{figure}

As we can see all the three cases behaves similarly. Actually we have extended the plot until redshift $z \simeq 12$ to notice appreciable differences among the three cases. In what follows, we combine both previous results to numerically reconstruct the potential as a function of the field $V(\phi)$ for our three cases. This is displayed in Fig. (\ref{vdfiabc}).
As we can see, for case (a) the potential has a stable minimum around $\phi \simeq 0.6 \, \phi_0 $ and start to increase slowly until today. The potential for case (b) shows a local minimum around  $\phi \simeq 0.5\, \phi_0 $ and its value is increasing as the field $\phi$ approach its current value. The case (c), which is the based on the Planck best fit, implies the potential is convex, with a maximum around $\phi \simeq 0.6\, \phi_0$ after that its value decreases as the field approach its current value.\\

As can be see in Fig. (\ref{vdfiabc}), the reconstructed potentials differs significantly depending on what set of values take the parameters $\omega_{0}-\omega_{a}$. This implies that our reconstruction scheme is able to cover a wide range of possible potentials, although we can not assert that it is comprehensive. For example, our analysis differs from that made in \cite{Scherrer:2015tra} where two extra conditions where added: that the total EoS were in the range $-1 \leq \omega \leq 1$ and also that the field be rolling downhill the potential (in that case the quintessence potential). Here we permit the parameters to take values such that the field go into the phantom regime $\omega < -1$, and also we do not restrict to fields rolling down but also we can have field rolling upward. This freedom enables us to describe a wide range of field potentials.\\ 

We would like to emphasize that the numerical reconstruction for $V(\phi)$ coming from the expression (\ref{eq:potential}) will represent in this case the appropriate scalar field potential for which the mimetic gravity will be in good agreement with the cosmological observations under the CPL parametrization, i.e., given a set of values fitted by the observations for $\omega_{0}$ and $\omega_{a}$, we constructed the potential $V(\phi)$. Reconstructions for the scalar field potential can be also found in the context of inflation, see for instance the Refs. \cite{rec1, rec2}.   

%%%%%%%%%%%%%%%%%%%%%%%%%%%%
\section{Final remarks}
\label{sec:final}
%%%%%%%%%%%%%%%%%%%%%%%%%%%%%
In this work we explored the Mimetic Gravity approach from a different perspective in order to describe the late times epoch of the observable Universe, i.e., we did not adopt any specific Ansatz for the scalar field potential, $V(\phi)$. As discussed before, this potential is the responsible of the dark energy behavior at late times for this kind of Universe. This can be seen in the expression (\ref{eq:multiplier}) for the Lagrange multiplier, which deviates from the typical $a^{-3}$ decay obtained in the mimetic description. The resulting constructed potential as a function of the cosmological redshift and its posterior numerical reconstruction as a function of the scalar field were possible with the assumption of the CPL parametrization for the parameter state of the fluid. In the discussion of our results we considered three cases labeled as (a), (b) and (c); which correspond to the best fit for the constants $\omega_{0}$ and $\omega_{a}$ of the CPL parametrization obtained with different sets of cosmological data, the case (c) was taken from the latest results of the Planck collaboration \cite{planck}. Then, in each case we obtain a different shape for the potential; according to its functional form, we will have that $V(z\rightarrow -1) \gg 1$ for $\omega_{a} > 0$ and $V(z\rightarrow -1)\rightarrow 0$ for $\omega_{a} < 0$. This latter case is present in the pair of values (c). On the other hand, from Eq. (\ref{eq:densityb}) we can see that the potential determines the behavior of the energy density $\rho_{\mathrm{MG}}$ and of course also the behavior of the Hubble parameter. Besides, for the cases (a) and (b) we have a future singularity at $z=-1$ for $E(z)$, therefore the model admits a little big rip.\\

It is worthy to mention that for the cases (a) and (b) we obtained desirable scenarios when the coincidence and deceleration parameters are computed. By assuming the standard form for the matter content and the dark energy sector described by mimetic gravity, we observed that the coincidence parameter is less than unity at present time and as we approach the far future this parameter decays to zero. This late behavior is not observed for the case (c), on contrary, the coincidence parameter starts to grow close to the far future. For the deceleration parameter we observe accelerated expansion for the cases (a) and (b) along cosmic evolution while in the case (c) the accelerated expansion is only a transient behavior, eventually the model has a smooth transition to a decelerated phase around $z \simeq -0.7$. In these scenarios, the Planck results are not favored by the nature of the observable Universe.    

\section*{Acknowledgments}
M. C. acknowledges support from S.N.I. (CONACyT-M\'exico). P. S. was supported in part by FONDECYT Grant No. 1180681 from the Government of Chile.

\end{document}